\input psfig.sty
\input lecproc.cmm
\pageno=0
\contributionnext{Primordial Heavy Element Production}
\author{T.\ Rauscher@1, 
F.-K. Thielemann@2}
\address{@1Institut f\"ur Kernchemie, Universit\"at Mainz, D-55099 Mainz, 
Germany
@2Institut f\"ur theoretische Physik, Universit\"at Basel, CH-4046 Basel, 
Switzerland}
\titlea{}{Introduction}
A number of possible mechanisms have been suggested
to generate density inhomogeneities in the early Universe 
which could survive until the onset of primordial 
nucleosynthesis (Malaney and Mathews 
1993). In this work
we are not concerned with how the inhomogeneities were
generated but we want to focus on the effect of such inhomogeneities on
primordial nucleosynthesis.
One of the proposed signatures of inhomogeneity, the 
synthesis of very heavy elements by neutron capture, was analyzed
for varying baryon to photon ratios $\eta$ and length scales $L$. 
A detailed discussion is published in (Rauscher et al.\ 1994b). 
Preliminary results 
can be found in (Thielemann et al.\ 1991; Rauscher et al.\ 1994a).
\titlea{}{Method}
After weak decoupling the vastly different mean free paths of protons 
and neutrons create a very proton rich environment in the initially high 
density regions, whereas the low density regions are almost entirely 
filled with diffused neutrons. Since the aim of the present 
investigation was to explore the production of heavy elements we 
considered only the neutron rich low density zones. High density, proton 
rich, environments might produce some intermediate elements via the 
triple-alpha-reaction but will in no case be able to produce heavy 
elements beyond iron. However, we included the effects of the 
diffusion of neutrons into the proton rich zones. Using a similar 
approach as introduced in (Applegate 1988; Applegate et al.\ 1988), 
the neutron diffusive loss 
rate $\kappa$ is given by
$$\kappa={4.2\times 10^4 \over 
(d/a)_{\rm cm\,MeV}}T_9^{5/4}(1+0.716T_9)^{1/2}\ts \hbox{\rm s}^{-1} 
\eqno(1)$$
in the temperature range $0.2<T_9<1$. We want to emphasize that this 
analytical treatment is comparable in accuracy to numerical methods 
using high resolution grids.
Thus, the only open parameter in 
the neutron loss due to diffusion is the comoving length scale of 
inhomogeneities ($d/a$). Small separation lengths between high density 
zones make the neutron
leakage out of the small low density zones most effective. Large
separation lengths make it negligible. (For a detailed 
derivation of (1), see also Rauscher et al.\ 1994b).

Our reaction network consists of two
parts, one part for light and intermediate nuclei ($Z\leq 36$), being 
a general nuclear network of 655 nuclei. 
The second part is an r-process
code (including fission)
extending up to $Z=114$ and containing all (6033) nuclei from the
line of stability to the neutron-drip line (see 
also Cowan et al.\ 1983).
These two networks were coupled together such that they both ran 
simultaneously
at  each time step, and  the number of neutrons produced and captured was
transmitted back and forth between them.
(For details of the included rates and new rate determinations see 
Rauscher et al.\ 1994b).
\titlea{}{Results and Discussion}
The most favorable condition for heavy element formation is an initial 
neutron abundance of $X_{\rm n}=1$ (i.e.\ only neutrons) in the low 
density region, leading to a density ratio
$\rho_{\rm low}/\rho_{\rm b}=1/8$ (Rauscher et al.\ 1994b). 
This leaves as 
open parameters the baryon to photon ratio 
$\eta=n_{\rm b}/n_{\gamma}=10^{-10}\eta_{10}$ and the comoving length scale 
($d/a$). Four sets of calculations have been performed, employing 
$\eta_{10}$ values of 416, 104, 52, and 10.4. Using the 
relation (B\"orner 1988)
$$\Omega_{\rm b} h_{50}^2=1.54 \times 10^{-2} (T_{\gamma o}/2.78{\rm K})^3
\eta_{10}\quad,\eqno(2)$$
with the present temperature of the microwave background 
$T_{\gamma o}$ and the Hubble constant $H_o=h_{50}\times 50$\ts km\ts 
s$^{-1}$\ts Mpc$^{-1}$,
this
corresponds to possible choices of ($h_{50},\Omega_{\rm b}$) being (2.5,1),
(1.3,1), (1,0.8), and (1,0.16).
The range covered in $\eta_{10}$
extends from roughly a factor of 2.2 below the lower limit to a factor 
of
13 above the upper limit for $\eta$ in the standard big bang. For each
of the $\eta$-values we considered four different cases of $d/a$:
(0) $\infty$ (negligible neutron back diffusion),
(1) 10$^{7.5}$ cm\ts MeV, (2) 10$^{6.5}$ cm\ts MeV, and (3) 10$^{5.5}$ 
cm\ts MeV.
(This corresponds to distances between nucleation sites of $\infty$, 
2700, 
270, and 27 m, respectively, at the time of the quark-hadron phase 
transition).

An exponential increase in r-process abundances with 
increasing $\eta$ was found. This is due to ``fission
cycling'', whereby each of the fission fragments
can form a fissionable nucleus again by neutron captures.
This is  of particular importance in environments with a long duration
of high neutron densities.
In an r-process with
fission cycling the production of heavy nuclei is not limited
to the r-process flow coming from 
light
nuclei but requires only a small amount of fissionable nuclei to be 
produced
initially. The total mass fraction of heavy nuclei is doubled with each 
fission cycle
and can thus be written as $X_{\rm r}= 2^n X_{\rm seed}$,
with $X_{\rm seed}$ denoting the 
initial
mass fraction of heavy nuclei. 
The cycle number $n$ is decreasing with decreasing neutron number 
density $n_{\rm n}$ (and increasing temperature $T$) because the
reaction flux experiences longer half-lives when the r-process path is
moving closer to stability.

Since the formation of heavy elements beyond Fe and Kr is a very 
sensitive measure of $\eta$ it can be used to provide an independent 
upper limit for the product $\Omega_{\rm b} H_0^2$. Figure 1 shows
observational (upper) limit on possible primordial 
heavy element abundances (Cowan et al.\ 1991; Beers et al.\ 1992; Mathews et al.\ 
1992) compared to our results. Also shown are the limits for 
$\Omega_{\rm b} H_0^2$ from comparison of observed
(Meyer et al.\ 1991; Kurki-Suonio et al.\ 1990; Ryan et al.\ 1992;
Duncan et al.\ 1992) and calculated light element abundances.
The tightest constraints are given by the light elements 
including Li, Be, and B (however, see recent doubts on the primordial
$^7$Li abundance in (Deliyannis et al.\ 1993)) 
for which the conditions cannot differ much from the standard big bang.

To study the influence of uncertainties in the reaction rates 
some test calculations were performed with a 
variation of the $^8$Li($\alpha$,n)$^{11}$B rate. Recent 
experiments (Mao et al.\ 1994) 
seem to suggest that the rate used in our calculations (Rauscher et al.\ 
1992) has to be 
increased by a factor of 3. 
Since the Li-rate is only affecting the seed abundances $X_{\rm seed}$, 
only a linear dependence of the resulting r-process abundances 
($X_{\rm r}=2^n X_{\rm seed}$) was found.
The same is true for the $^{18}$O(n,$\gamma$)$^{19}$O rate which 
was changed by a factor of 10 in a recent investigation (Beer et al.\ 
1994). 
The total change in heavy element abundances by a factor of 30
is also shown in Fig.\ 1. However, this underlines that not all 
reactions of importance are fully
explored, yet, and future changes can be expected.

Provided that density
fluctuations exist with large length scales compared to the neutron
diffusion length, the
limits for $\eta_{10}$ or $\Omega_{\rm b} h_{50}^2$ change to 104 and 1.6,
respectively, at which heavy element
abundances are produced in inhomogeneous big bang models at a level
comparable to the ones seen at lowest observable metallicities.
This reduces the difference between the constraints from light and heavy
elements, although the light element constraint is still tighter.

{\bf Acknowledgement:} TR is an Alexander 
von Humboldt fellow.

\begrefchapter{References}
\ref Applegate, J.H. (1988): Phys.\ Rep.\ {\bf 163} 141
\ref Applegate, J.H., Hogan, C.J., Scherrer, R.J. (1988): Ap.\ 
     J.\ {\bf 329} 572
\ref Beer, H., K\"appeler, F., Wiescher, M. (1994): in {\it Capture 
     Gamma-Ray Spectroscopy}, ed.\ by J. Kern (IOP, Bristol), p.\ 756
\ref Beers, T., Preston, G.W., Shectman, S.A. (1992): Astron.\ 
     J. {\bf 103} 1987
\ref B\"orner, G. (1988): {\it The Early Universe} (Springer, New York)
\ref Cowan, J.J., Cameron, A.G.W., Truran, J.W. (1983): Ap.\ 
     J. {\bf 265} 429
\ref Cowan, J.J., Thielemann, F.-K., Truran, J.W. (1991): 
     Phys.\ Rep.\ {\bf 208} 267
\ref Deliyannis, C.P., Pinsonneault, M.H., Duncan, D.K. (1993): 
     Ap.\ J. {\bf 414} 740
\ref Duncan, D.K., Lambert, D.L., Lemke, D. (1992): Ap.\ J. 
     {\bf 401} 584
\ref Kurki-Suonio, H., Matzner, R.A., Olive, K.A., Schramm, D.N. (1990):
     Ap.\ J. {\bf 353} 406
\ref Malaney, R.A., Mathews, G.J. (1993): Phys.\ Rep.\ {\bf 
     229} 145
\ref Mao, Z.Q., Vogelaar, R.B., Champagne, A.E., Blackmon, J.C.,
     Das, R.K., Hahn, K.I., Yuan, J. (1994): Nucl.\ Phys.\ {\bf A567} 125
\ref
\vskip 1 truecm
\psfig{file=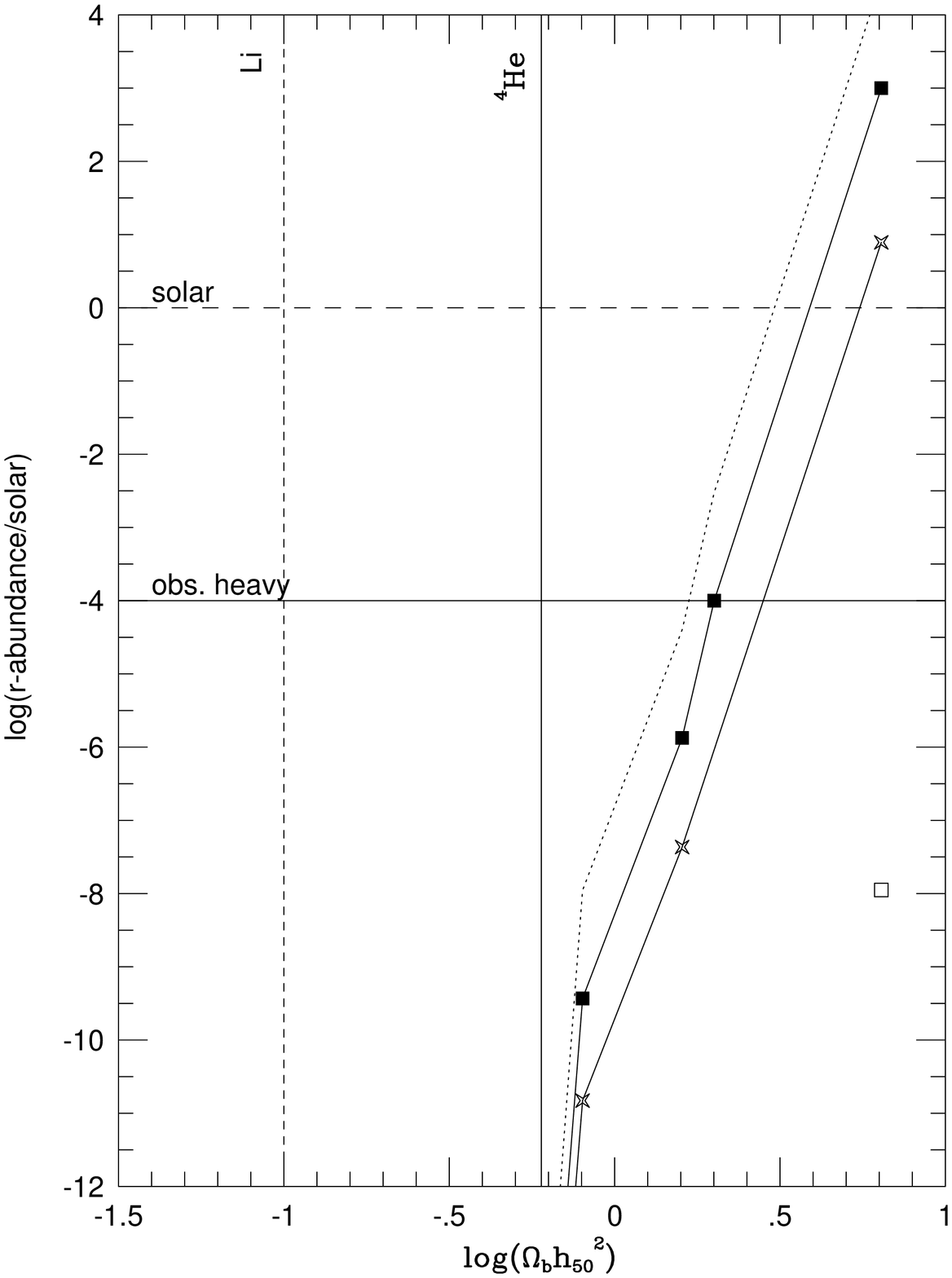,width=10.5cm,height=10.5cm}
\figure{1}{Limits on $\Omega_{\rm b} h_{50}^2$ from light and heavy element 
abundances. Shown are the results for 
cases 0 (full 
sq.), 1 (crosses), 2 (open sq.), and case 0 with enhanced rates (dotted). 
(The lines are drawn to guide the 
eye). 
The limits 
resulting from the
calculated values for the {\it light} elements are given by the 
vertical lines. (See text)}
\vskip 1 truecm
\ref Mathews, G.J., Bazan, G., Cowan, J.J. (1992): Ap.\ J. 
     {\bf 391} 719
\ref Meyer, B.S., Alcock, C.R., Mathews, G.J., Fuller, G.M. (1991):
     Phys.\ Rev.\ D {\bf 43} 1079
\ref Rauscher, T., Applegate, J.H., Cowan, J.J.,
     Thielemann, F.-K., Wiescher, M. (1994a): in Proc.\ Europ.\ Workshop on
     Heavy Element Nucleosynthesis, ed.\ by E. Somorjai, Z. F\"ul\"op 
     (Inst.\ Nucl.\ Res., Debrecen), p.\ 121
\ref Rauscher, T., Applegate, J.H., Cowan, J.J., 
     Thielemann, F.-K., Wiescher, M. (1994b): Ap.\ J. {\bf 429} 499
\ref Rauscher, T., Gr\"un, K., Krauss, H., Oberhummer, H., Kwasniewicz, 
     E. (1992): Phys.\ Rev.\ C {\bf 45} 1996
\ref Ryan, S.G., Norris, J.E., Bessel, M.S., Deliyannis, C.P. 
     (1992): Ap.\ J. {\bf 388} 184
\ref Thielemann, F.-K., Applegate, J.H., Cowan, J.J., 
     Wiescher, M. (1991): in {\it Nuclei in the Cosmos}, ed.\ by H. Oberhummer 
     (Springer, Heidelberg), p.\ 147
\endref
\byebye